\documentclass[aps,twocolumn,showpacs,prb,superscriptaddress]{revtex4-2}
\usepackage{mathtools}
\usepackage{times}
\usepackage{bm}
\usepackage{dsfont,amsthm,amsbsy}
\usepackage{verbatim}
\usepackage{amssymb}
\usepackage{amsmath}
\usepackage{bbm}
\usepackage{graphicx}
\usepackage{epstopdf}
\usepackage{subfigure}
\usepackage{natbib}
\usepackage{epsfig}
\usepackage{amsfonts}
\usepackage{mathrsfs}
\usepackage{sidecap}
\usepackage{lipsum}
\usepackage[toc,page,title,titletoc,header]{appendix}
\usepackage[colorlinks,linkcolor=blue,citecolor=blue,anchorcolor=blue,urlcolor=blue]{hyperref}
\usepackage{hyperref}
\usepackage{resizegather}
\usepackage{tikz}
\usepackage{float}
\usepackage{mathbbol}
\usepackage{cancel}
\usepackage{upgreek}

\usepackage{xcolor}
\colorlet{BLUE}{blue} 
\colorlet{blue}{black}
\colorlet{BLUE}{black} 

\makeatletter
\@clubpenalty=\clubpenalty   
\makeatother

\newcommand{\bl}{\begin{aligned}}
\newcommand{\el}{\end{aligned}}

\def\be{\begin{equation}}
\def\ee{\end{equation}}

\def\bi{\begin{itemize}}
\def\ei{\end{itemize}}
\def\bn{\begin{enumerate}}
\def\en{\end{enumerate}}
\def\bea{\begin{eqnarray}}
\def\eea{\end{eqnarray}}

\def\ba{\begin{array}}
\def\ea{\end{array}}
\def\bd{\begin{displaymath}}
\def\ed{\end{displaymath}}

\def\tr{{\rm tr}}

\begin{document}
\title{Comment on: "Scaling and Universality at Noisy Quench Dynamical Quantum Phase Transitions"}


%
\author{Jesko Sirker}
\affiliation{Department of Physics and Astronomy, University of Manitoba, Winnipeg R3T 2N2, Canada}
\affiliation{Manitoba Quantum Institute, University of Manitoba, Winnipeg R3T 2N2, Canada}

\date{\today}

\begin{abstract}
In Ref.~\cite{Ansari25}, dynamical quantum phase transitions (DQPTs)---non-analyticities in the Loschmidt return rate at critical times---are investigated in the presence of noise for a two-band model. The authors report that DQPTs persist even after averaging over the noise and they use their results to derive dynamical phase diagrams. \textcolor{blue}{The protocol used approximates the noise-averaged mixed state, obtained using a master equation, by a pure state, characterized by its excitation probability. In this comment we rigorously show that: (1) This approximation is exponentially poor in the thermodynamic limit. (2) When using the correct metric, the Loschmidt echo of two density matrices in any two-dimensional Hilbert space can become zero if and only if {\it both} density matrices are pure, ruling out DQPTs for non-zero noise. (3) An a posteriori reinterpretation of the results as an interferometric protocol is possible but such a protocol is unsuitable to investigate the effects of noise on DQPTs because it is inherently blind to decoherence.} We also investigate alternative natural ways to average over noise realizations and show that in all of them DQPTs are smoothed out.
\end{abstract}

\maketitle

\section{Introduction}
A significant portion of research on non-equilibrium dynamics has concentrated on the behavior of systems subjected to quenches and ramps. Here, an initial quantum state $\rho(0)$ is prepared and then, in the case of a quench, time-evolved with a time-independent Hamiltonian or, in the case of a ramp, with a Hamiltonian which has a time-dependent parameter. One way to study the ensuing dynamics is to consider the Loschmidt echo $|L(t)|^2$ which provides a measure on how far the time-evolved state $\rho(t)$ is from the initial state $\rho(0)$. Of particular interest are critical times $t_c$ where $L(t_c)=0$. These times define what in the literature has been called a DQPT \cite{HeylPolkovnikov,Andraschko2014,Heyl2018}.

\textcolor{blue}{The authors want to study the effect of noise on DQPTs and motivate their study by "Thus, understanding the effects of noise on Hamiltonian evolution is crucial for (i) accurately predicting experimental outcomes and (ii) designing advanced experimental setups that are resilient to noise". Among other results, the authors find that "In the presence of a Gaussian white noise, we find that this critical sweep velocity decreases by increasing the noise strength, and scales linearly with the square of the noise intensity." The method used consists of first solving a noise master equation which leads to a noise-averaged mixed state. This mixed state is then replaced by a pure state, characterized by its excitation probability ("we assumed the system’s state after the quench to be the noise-averaged pure state, ...; therefore, the reliability of our
results hinges on this assumption.") and calculating the Loschmidt echo for this pure state. The goal of this comment is to show that this pure-state approximation is exponentially poor and that any protocol which keeps only the excitation probability but neglects the coherences in the time-evolved density matrix is inherently blind to the most important aspect of noise: decoherence. We show that the effects of noise can be properly understood from the noise-averaged mixed state obtained from the master equation by employing the Uhlmann-Bures metric. Furthermore, we discuss alternative protocols for noise-sensitive Loschmidt echoes and show that in all of them DQPTs are smoothed out.}

In the most studied case, both the initial state $\rho(0)$ and the time-evolved state $\rho(t)$ are {\it pure states}. The Loschmidt amplitude is then simply defined by
\begin{equation}
    \label{L1}
    L(t)=\langle\Psi(0)|\Psi(t)\rangle
\end{equation}
with $\rho(t)=|\Psi(t)\rangle\langle\Psi(t)|$. If both states are instead impure, then the proper metric is the Uhlmann-Bures metric \cite{Uhlmann,Bures,SedlmayrFleischhauerSirker}
\begin{equation}
    \label{L2}
    \mathcal{L}(t)=\tr\sqrt{\sqrt{\rho(0)}\rho(t)\sqrt{\rho(0)}}
\end{equation}
which reduces to $|L(t)|$ with $L(t)$ defined in Eq.~\eqref{L1} if the density matrices $\rho$ are pure. Because overlaps scale exponentially with system size, it is typically not the Loschmidt echo which is considered but rather the Loschmidt return rate. For a one-dimensional system of size $N$, this return rate is defined by
\begin{equation}
    \label{return}
    g(t)=-\lim_{N\to\infty}\frac{1}{N}\ln|L(t)|^2,\,\mbox{or}\quad \mathcal{G}(t)=-\lim_{N\to\infty}\frac{1}{N}\ln|\mathcal{L}(t)|^2,
\end{equation}
respectively. In one dimension, zeroes of the Loschmidt echo manifest themselves as non-analyticities ('cusps') in the return rate.

\section{Considered setup}
The setup considered is shown in Fig.~1 of Ref.~\cite{Ansari25}. The system at the initial time $t_i$ is prepared in the ground state, a pure state, of a given Hamiltonian $H^i=H_0+h_i H_1$ with a magnetic field $h_i$. Then, in the time interval $[t_i,t_f)$ the system undergoes a linear ramp in the magnetic field $h_0(t)=h_i+\frac{h_f-h_i}{t_f-t_i}(t-t_i)$ so that at end of the ramp at time $t_f$---the authors choose $t_f=0$---the magnetic field is given by $h_f$. Accordingly, the ramp velocity is $v=(h_f-h_i)/(t_f-t_i)$. During the entire ramp, classical Gaussian white noise is supposed to be present which is described by the field $R(t)$ with correlation function $\langle R(t)R(t')\rangle = \xi^2\delta(t-t')$. The total field acting during the ramp is thus $h(t)=h_0(t)+R(t)$. For times $t\geq t_f$ the field remains at $h_f$ and there is no noise. Noise is present only during the ramp. The magnetic field $h(t)$ couples to the total staggered moment $H_1=\sum_n (-1)^n S^z_n$ in Ref.~\cite{Ansari25}. The following general arguments, however, are valid for any coupling. 

A step in Ref.~\cite{Ansari25} of crucial relevance for this comment is that at time $t=t_f=0$ the {\it noise-averaged density matrix} $\rho(t)$ is calculated using the master equation
\begin{equation}
    \label{rho_av}
    \dot{\rho}(t)=-i[H_{0}(t),\rho(t)]-\frac{\xi^{2}}{2}\Big[H_{1},\Big[H_{1},\rho(t)\Big]\Big].
\end{equation}
where $H_0(t)=H_0+h_0(t) H_1$ with $H_0$ being the considered time-independent Hamiltonian. This is a Lindblad master equation of dephasing type for the density matrix $\rho(t)$. The authors then want to consider the Loschmidt echo between the noise-averaged $\rho(t_f=0)$, obtained from Eq.~\eqref{rho_av}, and the state $\rho(t)$ for $t>0$ obtained by time-evolving $\rho(0)$ with the time-independent Hamiltonian $H^f=H_0+h_f H_1 $.

Surprisingly, the authors find that even after noise averaging over the initial state, the return rate shows non-analyticities (see Fig.~5 of Ref.~\cite{Ansari25}). I.e., DQPTs do still exist in the {\it noise-averaged} return rate. Based on these results, the authors also derive a dynamical phase diagram which consists of three phases (see Fig.~4 of Ref.~\cite{Ansari25}): a phase with no DQPTs and two phases with DQPTs. The phases with DQPTs are distinguished further by whether they show two critical momentum modes or multiple critical momentum modes (modes where the excitation probability is $1/2$). The latter phase, according to the phase diagram in Ref.~\cite{Ansari25}, only exists in the presence of noise. I.e. it is a novel dynamical phase created by noise.

\section{Two rigorous theorems ruling out the main findings of Ref.~\cite{Ansari25}}
In the following we prove two theorems which show that the main findings of Ref.~\cite{Ansari25} cannot hold \textcolor{blue}{for protocols sensitive to noise, i.e., protocols which take the coherences (off-diagonal elements) of $\rho$ into account such as the Uhlmann-Bures metric.}\\

{\bf Theorem 1:} Except for trivial cases, the solution of the Lindblad Master equation \eqref{rho_av} is a mixed state.\\
{\it Proof:} Consider the purity $P=\tr\rho^2(t)$. Then
\begin{eqnarray}
    \label{purity}
    \dot P(t) &=& 2\tr(\rho\dot\rho)=2\tr\left(\rho\left\{-i[H_0,\rho]-\frac{\xi^2}{2}[H_1,[H_1,\rho]]\right\}\right) \nonumber \\
    &=& -\xi^2 \tr\left(\rho[H_1,[H_1,\rho]]\right)
    \end{eqnarray}
where we have used $\tr(\rho[H_0,\rho])=0$ because of the cyclic invariance of the trace and, for brevity, have written $\rho=\rho(t)$ and $H_0=H_0(t)$. We can now use that for Hermitian operators $A,B$ the identity $\tr(B[A,[A,B]])=\tr([A,B]^\dagger [A,B])\geq 0$ holds which is straightforward to prove using again the cyclic invariance of the trace. We therefore find for the purity
\begin{equation}
    \label{purity2}
    \dot P(t) = -\xi^2 \tr\left([H_1,\rho]^\dagger [H_1,\rho]\right)\leq 0
\end{equation}
which means that the purity is a non-increasing function. If we start from a pure state then the state remains pure if and only if $[H_1,\rho(t)]=0$ for all times $t$. In this case we are in a decoherence-free subspace and the dissipator vanishes identically leaving only the unitary dynamics $\dot\rho =-[H_0,\rho]$. A typical example of such decoherence-free dynamics is the case $[H_0,H_1]=0$ where we start in an eigenstate of $H_0$ which then in turn is also an eigenstate of $H_1$. If we want to study the effects of noise on DQPTs then we are not interested in cases where the noise is inactive and the dynamics decoherence free. Indeed, in the example considered in Ref.~\cite{Ansari25}, $H_1$ is the staggered magnetic field and $[H_0,H_1]\neq 0$. The purity is therefore a monotonically decreasing function and the fixed point of the Lindblad Master equation \eqref{rho_av} is the maximally mixed state.\\

{\bf Theorem 2:} In a two-dimensional Hilbert space, the Loschmidt echo $\mathcal{L}(t)$ can have zeroes if and only if {\it both} the initial state $\rho(0)$ and the final state $\rho(t)$ are pure. In this case $\mathcal{L}(t)=|L(t)|=|\langle\Psi(0)|\Psi(t)\rangle|$ and zeroes occur if critical times $t_c$ exist where $|\Psi(0)\rangle$ and $|\Psi(t)\rangle$ are orthogonal.\\
{\it Proof:} Since $M(t) = \sqrt{\rho(0)}\rho(t)\sqrt{\rho(0)}$ is positive semidefinite, this implies
\begin{equation}
    \label{T2_1}
   \mathcal{L}(t)=\tr\sqrt{M(t)}=0 \Longleftrightarrow M(t)=0 \, .
\end{equation}
Furthermore,
\begin{equation}
    \label{T2_2}
    M(t) = 0  \Longleftrightarrow \mbox{supp}[\rho(0)]\perp \mbox{supp}[\rho(t)],
\end{equation}
i.e., $M(t)$ vanishes if and only if the supports of the two density matrices are orthogonal to each other. This result is general and valid in any dimension. 

In two-dimensional Hilbert spaces, the possible ranks of each of the two density matrices are $1$ or $2$. If either of the density matrices has rank $2$ then their support is the entire two-dimensional Hilbert space and there is thus no orthogonal complement. It follows that $\mathcal{L}(t)=0$ is possible if and only if both density matrices have rank $1$ which means they are pure. In this case $\rho(0)=|\Psi(0)\rangle\langle\Psi(0)|$ and $\rho(t)=|\Psi(t)\rangle\langle\Psi(t)|$ implying that $\mathcal{L}(t)=|\langle\Psi(0)|\Psi(t)\rangle| = |L(t)|$, see Eq.~\eqref{L1} and Eq.~\eqref{L2}, and $L(t)=0$ if the two pure states $|\Psi(0)\rangle$ and $|\Psi(t)\rangle$ are orthogonal to each other. We note that in Hilbert spaces with dimensions larger than $2$ it is possible for the Loschmidt echo to vanish also for impure states. For example, in $d=4$ we can have $\rho(0)=\frac{1}{2}|0\rangle\langle 0| + \frac{1}{2}|1\rangle\langle 1|$ and $\rho(t)=\frac{1}{2}|2\rangle\langle 2| + \frac{1}{2}|3\rangle\langle 3|$ which have orthogonal support. Note, however, that even in higher dimensions this is not a generic scenario and will require some fine tuning.  

In Ref.~\cite{Ansari25} a translationally invariant, one-dimensional, two-band model is considered. As a consequence, the Hilbert space factorizes into two-dimensional Hilbert spaces for each momentum mode,
in particular, 
\begin{equation}
    \label{DM}
    \rho(t) = \bigotimes_k \rho_k(t)
\end{equation}
where $\rho_k(t)$ are $2\times 2$ matrices. It follows that 
\begin{equation}
    \label{DM2}
    \mathcal{L}(t)=\prod_k \mathcal{L}_k(t)=\prod_k \tr\sqrt{\sqrt{\rho_k(0)}\rho_k(t)\sqrt{\rho_k(0)}}
\end{equation}
and $\mathcal{L}(t)=0$ if and only if at least one momentum $k$ exists with $\mathcal{L}_k(t)=0$. As we have shown in Theorem 2 this can happen if and only if both $\rho_k(0)$ and $\rho_k(t)$, which live in a two-dimensional Hilbert space, are pure. However, in Theorem 1 we have proven that the density  matrix $\rho_k(0)$---obtained after the noisy ramp by taking the noise average---is an impure state. We have thus rigorously proven that there are no DQPTs under the noise-averaging procedure considered in Ref.~\cite{Ansari25} for any non-zero noise amplitude. For any noise amplitude $\xi\neq 0$ there is only a 'no-DQPT' phase, and not three phases including two with DQPTs as shown incorrectly in Fig.~4 of Ref.~\cite{Ansari25}.

As a further check of Theorem 2 and as a tool useful to calculate the Loschmidt echo $\mathcal{L}_k(t)$ we note that in two dimensions we can write $\rho_k(t)=\frac{1}{2}(I+\bm{r}_k(t) \cdot \bm{\sigma})$ where $\bm{\sigma}$ is the vector of Pauli matrices and $I$ is the $2\times 2$ identity. From this representation it follows that
\begin{equation}
    \label{DM3}
    \mathcal{L}_k(t) =\sqrt{\frac{1}{2}\left(1+\bm{r}_k(0)\cdot\bm{r}_k(t)+\sqrt{(1-\|\bm{r}_k(0)\|^2)(1-\|\bm{r}_k(t)\|^2)}\right)} \, .
\end{equation}
This is just the standard expression for the Uhlmann fidelity of two qubit density matrices. From this formula it also immediately follows that $\mathcal{L}_k(t)=0$ if and only if $\|\bm{r}_k(0)\|=\|\bm{r}_k(t)\|=1$ (both states are pure) and $\bm{r}_k(0)\cdot\bm{r}_k(t)=-1$ (they are orthogonal).

To summarize, we have rigorously proven two theorems which are of general use when considering the Loschmidt echo in the presence of dissipation and which, more specifically, directly rule out the main findings of Ref.~\cite{Ansari25} \textcolor{blue}{if the proper Uhlmann-Bures metric, which is sensitive to coherences, is used}.

\section{\protect\textcolor{blue}{Pure state approximations and interferometric protocols}}
Next, we will discuss the concrete issues leading to the incorrect dynamical phase diagram in Ref.~\cite{Ansari25} \textcolor{blue}{which result from "we assumed the system’s state after the quench to be the noise-averaged pure state", i.e., approximating the mixed state obtained from the noise master equation \eqref{rho_av} by a pure state}. The issues are general and we do not need to consider the concrete form of the Hamiltonian used in Ref.~\cite{Ansari25}. 

We can write an initial pure state for a translationally invariant model as $|\Psi(0)\rangle=\bigotimes_k |\Psi_k(0)\rangle$ and the Hamiltonian as $H=\bigoplus_k H_k$. Furthermore, for a two-band model as considered in Ref.~\cite{Ansari25} there are only two states for each $k$-mode so we can write $|\Psi_k(0)\rangle=v_k|\alpha_k\rangle +u_k|\beta_k\rangle$ with $|v_k|^2+|u_k|^2=1$. We can also assume that $H_k|\alpha_k\rangle=-\varepsilon_k|\alpha_k\rangle$ and $H_k|\beta_k\rangle=\varepsilon_k|\beta_k\rangle$. The {\it pure-state Loschmidt echo} is then given by $L(t)=\prod_k L_k(t)$ with
\begin{eqnarray}
    \label{twoband1}
    L_k(t) &=& |\langle\Psi_k(0)|e^{-iH_k t}|\Psi_k(0)\rangle|^2\nonumber \\
     &=& 1-4p_k(1-p_k)\sin^2(\varepsilon_k t) 
\end{eqnarray}
where we have defined $p_k=|u_k|^2=|\langle\Psi_k(0)|\beta_k\rangle|^2$. This formula for the Loschmidt echo is correct for two-band models if (i) the initial state $|\Psi_k(0)\rangle$ is pure, and (ii) we time evolve the system with a time-independent Hamiltonian $H_k$. 

If we calculate, as the authors do, the probability
\begin{equation}
    \label{pk}
    \tilde p_k =\langle\beta_k|\rho_k(0)|\beta_k\rangle
\end{equation}
then this can indeed be interpreted as the probability of the mixed state $\rho_k(0)$ collapsing onto the pure state $|\beta_k\rangle$ in a measurement. However, this probability is not the pure-state amplitude $p_k$ which enters into the Loschmidt echo formula \eqref{twoband1}, $\tilde p_k \neq p_k$. The authors assume that these two probabilities are the same, \textcolor{blue}{which appears to be the central approximation of the paper.} To see whether \textcolor{blue}{this approximation is valid}, we start from the general form of a $2\times 2$ density matrix $\rho_k(0)=\frac{1}{2}(I+\bm{r}\cdot\bm{\sigma})$ where $I$ is the $2\times 2$ identity matrix. This form of $\rho_k(0)$ automatically fulfills $\tr\rho_k(0)=1$ and also guarantees that $\rho_k(0)$ is hermitian. The condition that a density matrix has to be positive semi-definite further implies $\|\bm{r}\|\leq 1$. The pure states are lying on the surface of the Bloch sphere, i.e.~for pure states we have $\|\bm{r}\|=1$. W.l.o.g.~we can assume, by an appropriate basis choice in the two-level space, that $\beta_k=(1\; 0)^T$. Then $\tilde p_k=\langle\beta_k|\rho_k(0)|\beta_k\rangle=(1+r_z)/2$. If we set $\tilde p_k= p_k$ then we assume that we can replace the mixed state $\rho_k(0)$ by the pure state $|\Psi_k(0)\rangle = \sqrt{(1-r_z)/2}|\alpha_k\rangle + \sqrt{(1+r_z)/2} |\beta_k\rangle$. In other words, the assumption is that the density matrix $\rho_k(0)$ is well approximated by the pure-state density matrix $\bar{\rho}_k=\frac{1}{2}(I+\bm{s}\cdot\bm{\sigma})$ with $\bm{s}=(\sqrt{1-r_z^2},0,r_z)$. This is the pure state obtained by keeping the $z$ component the same and maximizing the coherence in the $x$ component. However, this is {\it not} the pure state which best approximates the general mixed state $\rho_k(0)$. The best approximation is instead obtained by moving along the direction of the $\bm{r}$ vector to the surface of the Bloch sphere, i.e.~the best approximation is the pure state $\tilde\rho_k(0)=\frac{1}{2}(I+\hat{\bm{r}}\cdot\bm{\sigma})$ with $\hat{\bm{r}}=\bm{r}/\|\bm{r}\|$. We can see this by considering the mixed-state fidelity
\begin{equation}
    \label{fidelity}
F(\rho_1,\rho_2)=\left(\tr\sqrt{\sqrt{\rho_1}\rho_2\sqrt{\rho_1}}\right)^2 \, .
\end{equation}
For the approximation chosen in Ref.~\cite{Ansari25} we find $\bar F_k(\rho_k,\bar\rho_k)=(1+\bm{r}\cdot\bm{s})/2=(1+r_z^2+r_x\sqrt{1-r_z^2})/2$ whereas $\tilde F_k(\rho_k,\tilde\rho_k)=(1+\|\bm{r}\|)/2$ which is the maximal fidelity possible. We therefore see that the construction used in the paper only reproduces an optimal pure-state approximation of the mixed state $\rho_k(0)=(I+\bm{r}\cdot\bm{\sigma})$ if $\bm{r}=(r_x,0,0)$, i.e., if the $\bm{r}$-vector points in $x$ direction. 

More fundamentally, even if we choose the optimal pure-state approximation, we have $\tilde F_k<1$ so that any such approximation is exponentially poor in system size
\begin{equation}
    \label{fidelity2}
    F=\prod_k F_k \sim \mathcal{O}(\text{e}^{-N}) \, .
\end{equation}
This can be most clearly seen if we consider the fixed point of the Lindblad Master equation \eqref{rho_av} which is the completely mixed state $\rho^{\rm fp}_k=I/2$. In this case any pure state is an equally poor approximation with fidelity $F^{\rm fp}_k=1/2$ leading to a total fidelity $F=2^{-N}$. 

\textcolor{blue}{The discussion so far takes the construction of Ref.~\cite{Ansari25} at face value as an approximation of the mixed state $\rho_k(0)$ by a pure state, which is precisely how the authors describe their procedure (see Sec.~II). One might nevertheless attempt to defend the result by reinterpreting {\it a posteriori} the rate function not as a pure-state approximation but as an {\it exact} quantity for the mixed state---for instance the interferometric (Pancharatnam) Loschmidt echo $\mathcal{L}^{\rm{int}}_k(t)=\tr[\rho_k(0)\,e^{-iH_k t}]$, which has been employed for mixed states elsewhere \cite{HeylBudich_mixed,Bhattacharya_mixed}. We stress first that this is {\it not} the procedure described in Ref.~\cite{Ansari25}, whose authors explicitly evaluate the dynamical free energy under the assumption of a pure post-quench state, as quoted in Sec.~II. More importantly, such a reinterpretation does not resolve the question of noise either, for the following reason. Writing $H_k=\varepsilon_k\,\hat{\bm{n}}\cdot\bm{\sigma}$, where $|\beta_k\rangle$ is the $+\hat{\bm{n}}$ eigenstate, and $\rho_k(0)=\frac{1}{2}(I+\bm{r}\cdot\bm{\sigma})$, a direct calculation gives
\begin{eqnarray}
    \label{interf}    |\mathcal{L}^{\rm{int}}_k(t)|^2&=&\cos^2(\varepsilon_k t)+(\bm{r}\cdot\hat{\bm{n}})^2\sin^2(\varepsilon_k t) \nonumber \\
    &=& 1-4\tilde p_k(1-\tilde p_k)\sin^2(\varepsilon_k t),
\end{eqnarray}
with $\tilde p_k=\langle\beta_k|\rho_k(0)|\beta_k\rangle=(1+\bm{r}\cdot\hat{\bm{n}})/2$ as in Eq.~\eqref{pk}. This reproduces precisely the noisy return rate of Ref.~\cite{Ansari25} [their Eqs.~(2) and (14)]. Unlike the Uhlmann-Bures echo \eqref{L2}, this interferometric quantity can indeed vanish for mixed states. However, it depends on $\rho_k(0)$ {\it only} through the single projection $\bm{r}\cdot\hat{\bm{n}}$ and is completely independent of the length $\|\bm{r}\|$ of the Bloch vector---the coherences multiply the vanishing off-diagonal entries of the diagonal evolution operator and drop out of the trace. It therefore cannot resolve the purity of the state: a pure state and a mixed state sharing the same value of $\bm{r}\cdot\hat{\bm{n}}$ yield identical return rates for all times $t$. Since the entire physical effect of the noise is the reduction of the purity, i.e.~the shrinking of $\|\bm{r}\|$ (cf.~Theorem 1), a diagnostic that discards $\|\bm{r}\|$ carries no information whatsoever about the decoherence caused by the noise. The cusps it displays are properties of the population $\tilde p_k$ alone---which can be produced by entirely unrelated physical processes---and are not evidence that DQPTs survive noise. Thus, whether the construction of Ref.~\cite{Ansari25} is read as a pure-state approximation or as an exact interferometric quantity, it does not address the question the paper sets out to investigate.}

\section{Noise averaging always destroys DQPTs}
Having shown that the approach used in Ref.~\cite{Ansari25} is \textcolor{blue}{insensitive to noise}, we now discuss three inequivalent and natural ways proper noise averages can be obtained. In this section we will denote quantities $X$ which are noise averaged by $\bar{\bar X}=\mathbbm{E}_R[X]$. \\

{\bf Case 1:} As in the paper, the Master equation \eqref{rho_av} is used to obtain the noise-averaged initial mixed state $\bar{\bar{\rho}}(0)$. Using the Loschmidt echo $\mathcal{L}[\bar{\bar\rho}](t)$ based on the noise-averaged density matrix we obtain the return rate
\begin{equation}
    \label{g1}
    \bar{\bar{\mathcal{G}}}(t) = -\lim_{N\to\infty}\frac{1}{N}\ln|\mathcal{L}[\bar{\bar\rho}](t)|^2 \, .
\end{equation}

{\bf Case 2:} For each noise realization $n$, we obtain a new initial pure state $|\Psi_n(0)\rangle$. We then use the pure-state Loschmidt echo $|L_n(t)|^2 = |\langle\Psi_n(0)|\Psi_n(t)\rangle|^2$ and average over the noise realizations to obtain 
\begin{eqnarray}
    \label{g2}
  |\bar{\bar{L}}(t)|^2 &=& \lim_{M\to\infty} \frac{1}{M}\sum_{n=1}^M |L_n(t)|^2, \nonumber \\ 
  \bar{\bar g}(t) &=& -\lim_{N\to\infty}\frac{1}{N}\ln |\bar{\bar{L}}(t)|^2  \, .
\end{eqnarray}

{\bf Case 3:} Similar to case 2, we start from the pure state $|\Psi_n(0)\rangle$ but calculate the return rate for each realization first and only take the noise average at the very end
\begin{eqnarray}
    \label{g3}
    g_n(t) &=& -\lim_{N\to\infty}\frac{1}{N} \ln |L_n(t)|^2 \nonumber \\
  \bar{\bar g}'(t) &=& \lim_{M\to\infty}\frac{1}{M} g_n(t)  \, .
\end{eqnarray}

Case 1 is the noise average the authors of Ref.~\cite{Ansari25} set out to consider \textcolor{blue}{before invoking the pure state approximation}. We have rigorously proven through Theorems 1 and 2 that for two-band models no DQPTs exist for any non-zero noise. In case 2, each realization can have critical times $\tau_{n,k}$ ($k=1,\cdots$) such that $L_n(\tau_{n,k})=0$. However, the probability that a finite fraction of realizations has the exact same critical time is zero, because the critical times $\tau_{n,k}$ are continuous random variables that depend sensitively on the noise realization. Thus, $|\bar{\bar L}(t)|^2\neq 0$ for all times $t$ with probability $1$ and $\bar{\bar g}(t)$ is smooth. A similar argument also holds for case 3: Each individual return rate $g_n(t)$ can have cusps at times $\tau_{n,k}$, however, these non-analyticities occur at different random times $\tau_{n,k}$ and thus do not add up coherently. As a result, $\bar{\bar g}'(t)$ is again a smooth function.

\section{Conclusions}
Ref.~\cite{Ansari25} sets out to study the effect of noise on DQPTs---non-analyticities of the Loschmidt return rate at critical times. We have shown here that the authors' approach \textcolor{blue}{of replacing the mixed state by a pure state is insensitive to the main effect of noise, decoherence, and} therefore does not reflect the actual physics of the problem. \textcolor{blue}{This fundamental issue remains even if the results, a posteriori, are interpreted not as an approximation scheme but as an exact interferometric protocol.}

For the noise-averaging protocol considered in the paper---where the initial state after the ramp is obtained by averaging over the noise---we have rigorously proven two general theorems which demonstrate that a properly chosen Loschmidt return rate, \textcolor{blue}{which is sensitive to noise}, will never show DQPTs for non-zero noise levels in two-band models. We note that the second theorem appears to be relevant beyond the purview of this comment by showing that in $n$-band models with $n>2$ DQPTs are possible even for mixed states although this is expected to happen only in some fine-tuned models and not in generic noisy or finite-temperature systems.

In addition, we have also considered two other scenarios to obtain noise-averaged return rates. In these cases we are always working with pure states and individual noise realizations can show DQPTs because they correspond to unitary evolutions of pure states. However, the times at which DQPTs do occur are not coherent between different realizations. As a result, a noise averaging will always smooth out DQPTs also in these alternative approaches.

\textcolor{blue}{Overall, we have shown that the chosen protocol cannot answer the question how noise affects DQPTs because it is insensitive to decoherence and have discussed alternative protocols where decoherence is taken into account. As might have been expected on general physical grounds, averaging over noise then always does smooth out dynamical quantum phase transitions showing that the sharp transitions observed in Ref.~\cite{Ansari25} are artefacts of an unsuitable protocol.}

\acknowledgments
The author acknowledges support by the National Science and Engineering Research Council of Canada (NSERC) via the Discovery Grants program.

%

\end{document}